\documentclass[prd,aps,twocolumn,preprintnumbers, showpacs, nofootinbib,superscriptaddress,notitlepage]{revtex4-1}
\usepackage{amssymb,amsmath,graphicx,hyperref}
\usepackage{mathrsfs}
\usepackage{amsfonts}
\usepackage{amsmath}
\usepackage{slashed}
\usepackage{array}
\usepackage{verbatim}
\usepackage{epsfig}
\usepackage{graphicx}
\usepackage{color}
\usepackage[dvipsnames]{xcolor}

\newcommand{\beq}{\begin{eqnarray}}
\newcommand{\eeq}{\end{eqnarray}}

\newcommand{\nn}{\nonumber} 
\newcommand{\back}{\!\!\!\!}

\DeclareMathOperator\Det{Det}
\DeclareMathOperator\re{Re}
\DeclareMathOperator\im{Im}

\begin{document}

\title{Parton Distribution Function with Nonperturbative Renormalization\\ from Lattice QCD\\
}

\collaboration{Lattice Parton Physics Project (LP3)}

\author{Jiunn-Wei Chen}
\affiliation{
Department of Physics, Center for Theoretical Sciences, and Leung Center for Cosmology and Particle Astrophysics, National Taiwan University, Taipei, Taiwan 106}

\author{Tomomi Ishikawa}
\affiliation{T.~D.~Lee Institute, Shanghai Jiao Tong University, Shanghai, 200240, P. R. China}

\author{Luchang Jin}
\affiliation{Physics Department, Brookhaven National Laboratory, Upton, New York 11973, USA}

\author{Huey-Wen Lin}
\affiliation{Department of Physics and Astronomy, Michigan State University, East Lansing, MI 48824}
\affiliation{Department of Computational Mathematics, Michigan State University, East Lansing, MI 48824}

\author{Yi-Bo Yang}
\email{yangyibo@pa.msu.edu}
\affiliation{Department of Physics and Astronomy, Michigan State University, East Lansing, MI 48824}

\author{Jian-Hui Zhang}
\email{jianhui.zhang@ur.de}
\affiliation{Institut f\"{u}r Theoretische Physik, Universit\"{a}t Regensburg, D-93040 Regensburg, Germany}

\author{Yong Zhao}
\affiliation{Center for Theoretical Physics, Massachusetts Institute of Technology, Cambridge, MA 02139, USA
}

\preprint{MSUHEP-17-007,MIT-CTP/4912}

\begin{abstract}
We present lattice results for the isovector unpolarized parton
distribution with nonperturbative RI/MOM-scheme renormalization on the
lattice. In the framework of large-momentum effective field theory (LaMET), the full Bjorken-$x$ dependence of a momentum-dependent quasi-distribution is calculated on the lattice and matched to the ordinary lightcone parton distribution at one-loop order, with power corrections included. The important step of RI/MOM renormalization that connects the lattice and continuum matrix elements is detailed in this paper. A few consequences of the results are also addressed here.
\end{abstract}

\maketitle

\section{Introduction}
\label{sec:intro}

Parton distribution functions (PDFs) are probability densities of quarks and gluons seen by an observer moving at the speed of light relative to the hadron. They are universal nonperturbative properties of the hadron. In a global analysis the hard-scattering cross sections can be factorized into the PDFs and the short-distance matrix elements calculable in perturbation theory. Once known, the PDFs can be used as inputs to predict cross sections in high-energy scattering experiments, one of the major successes of QCD. Today, multiple collaborations provide regular updates concerning the phenomenological determination of the PDFs~\cite{Ball:2012cx,Ball:2014uwa,Harland-Lang:2014zoa,Dulat:2015mca,Alekhin:2017kpj,Owens:2012bv} using the latest experimental results, either focusing on medium-energy QCD experiments or high-energy ones such as those at the LHC. After the past half century of theoretical and experimental efforts, the precision needed in PDFs to further test the Standard Model has increased significantly, and experiments are planned to push further into unexplored or less known regions, such as sea-quark and gluonic structure. 

In this work, we continue the first-principles calculation of the PDFs using lattice QCD. 
In parton physics, the PDFs are defined as the nucleon matrix elements of quark or gluon correlation operators along the lightcone direction. For example, the unpolarized quark distribution is defined as
\begin{align}\label{eq:pdf}
q(x,\mu) \equiv \!\int\!\! \frac{d\xi^-}{4\pi} \, e^{-ixP^+\xi^-} 
\! \big\langle P \big| \bar{\psi}(\xi^-) \gamma^+ 
W(\xi^-\!,0)
\psi(0) \big|P\big\rangle ,
\end{align}
where $\mu$ is the renormalization scale in a given renormalization scheme, such as the $\overline{\text{MS}}$ scheme, the nucleon momentum $P_\mu=(P_0,0,0,P_z)$, $\xi^\pm = (t\pm z)/\sqrt{2}$ are the lightcone coordinates, and the Wilson line
\begin{align}
W(\xi^-,0) &= \exp\bigg(-ig \int_0^{\xi^-}d\eta^- A^+(\eta^-) \bigg) \,
\end{align}
is inserted to ensure gauge invariance. The main obstacle in directly computing PDFs from lattice QCD is their lightcone dependence. Since lattice QCD is formulated in Euclidean space which maps the whole Minkowski lightcone to a single point, the $\xi^-$ dependence is completely lost.
Early lattice studies of PDFs used the operator product expansion (OPE) to calculate their moments, which are matrix elements of local gauge-invariant operators~\cite{Detmold:2001dv,Detmold:2002nf,Dolgov:2002zm,Gockeler:2004wp}.
However, discretization error and operator mixing due to the breaking of rotational symmetry on the lattice make it hard to go beyond the first few moments.
There exist proposals for obtaining higher moments by using smeared sources~\cite{Davoudi:2012ya} or computing current-current correlators in Euclidean space~\cite{Liu:1993cv,Detmold:2005gg,Braun:2007wv,Chambers:2017dov}.
Recently, there is a lattice study on the feasibility of computing pion DA with Euclidean current correlators~\cite{Bali:2017gfr}.

Recently, Ji~\cite{Ji:2013dva,Ji:2014gla} proposed a new approach for the direct computation of parton physics, large-momentum effective field theory (LaMET). According to this approach, in order to get the normal PDF, one can start by calculating a ``quasi-PDF'', which is defined as a spatial correlation of partons along, say the $z$ direction, in a moving nucleon, 
\begin{align} \label{eq:quasipdf}
& \tilde{q}(x, P_z, \tilde{\mu}) 
= \int_{-\infty}^\infty \frac{dz}{4\pi}\ e^{ixP_zz} {\tilde{h}}(z,P_z,\tilde{\mu}) 
\,,
\\
& {\tilde{h}}(z,P_z,\tilde{\mu}) 
= {\big \langle P| O_{\Gamma} | P \big\rangle}
\,, \nn
\end{align} 
where {$O_{\Gamma}=\bar{\psi}(z) {\Gamma} W_z(z,0) \psi(0)$ with $\Gamma=\gamma_z$ or $\frac{P_z}{P_t}\gamma_t$}~{\cite{Xiong:2013bka,Radyushkin:2016hsy,Radyushkin:2017gjd,Radyushkin:2017cyf}}, $\tilde{\mu}$ is the renormalization scale in a particular scheme, and the spacelike Wilson line is
\begin{align}
W_z(z,0) &=\exp\left(ig\int_0^z dz' A^z(z') \right) \,.
\end{align} 
Unlike the definition in Eq.~\ref{eq:pdf}, which is invariant under a Lorentz boost along the $z$ direction, 
the quasi-PDF changes dynamically under such a boost and depends nontrivially on the nucleon momentum $P_z$. For a nucleon of mass $M_N$ moving with finite but large momentum $P_z\gg M_N, \Lambda_\text{QCD}$, LaMET allows us to match the quasi-PDF to the PDF through a factorization formula~\cite{Ji:2013dva,Ji:2014gla}:
\begin{align} \label{eq:factorization}
\tilde{q}(x, P_z, \tilde{\mu}) &= \int_{-1}^{+1} \frac{dy}{|y|} \ C\left(\frac{x}{y}, \frac{\tilde{\mu}}{P_z},\frac{\mu}{yP_z}\right) q(y,\mu)
\nn\\ 
&\quad
+ {\cal O}\bigg(\frac{M_N^2}{P_z^2}, \frac{\Lambda_{\text{QCD}}^2}{P_z^2} \bigg)
\ ,
\end{align}
where $C$ is the matching kernel, and the ${\cal O}(M_N^2/ P_z^2, \Lambda_{\text{QCD}}^2/ P_z^2)$ terms are power corrections suppressed by the nucleon momentum. Here $q(y,\mu)$ for negative $y$ corresponds to the antiquark contribution. The $\tilde{q}$ and $q$ have the same infrared (IR) divergences,
so the matching kernel $C$ depends on ultraviolet (UV) physics only and, thus, can be calculated in perturbative QCD.

There has been rapid development following Ji's proposal. 
Lattice-QCD calculations of the proton isovector quark distribution $f_{u-d}$~\cite{Lin:2014zya,Alexandrou:2015rja,Chen:2016utp,Alexandrou:2016jqi}, including the unpolarized, polarized and transversity cases, 
as well as the pion distribution amplitude~\cite{Zhang:2017bzy},
have been carried out within the LaMET approach. 
The one-loop matching kernels were calculated in the continuum theory for the isovector quark distributions in a transverse-momentum cutoff scheme in Ref.~\cite{Xiong:2013bka} and reproduced in Refs.~\cite{Ma:2014jla,Alexandrou:2015rja}; the matching for GPDs was addressed in Refs.~\cite{Ji:2015qla,Xiong:2015nua}. Recently also studies in lattice perturbation theory are available~\cite{Carlson:2017gpk,Briceno:2017cpo,Xiong:2017jtn}. The nucleon-mass corrections to all orders in
$M_N^2/P_z^2$ have already been derived in Refs.~\cite{Lin:2014zya,Chen:2016utp}
and included in the lattice calculations~\cite{Lin:2014zya,
Chen:2016utp}, while the higher-twist $O(\Lambda^2_\text{QCD}/P_z^2)$ correction was numerically removed by fitting the results at different $P_z$ with a polynomial of $1/P_z^2$ and extrapolating to infinite momentum~\cite{Lin:2014zya,Chen:2016utp}.

Despite many promising features in the previous lattice calculation of PDFs~\cite{Lin:2014zya,Alexandrou:2015rja,Chen:2016utp,Alexandrou:2016jqi,Zhang:2017bzy}, {one important
piece was missing until recently to form a complete image}: the lattice renormalization of the quasi-PDFs. 
The UV transverse-momentum cutoff scheme used in the one-loop matching computation~\cite{Xiong:2013bka,Ma:2014jla} is not the same regularization as used on the discretized lattice. To reduce systematic uncertainties from this mismatch, a proper renormalization of the bare lattice matrix elements is required. An alternative approach is to replace the lattice regularization by the gradient flow and match the continuum extrapolated results to the $\overline{\text{MS}}$ PDF~\cite{Monahan:2016bvm}, where the latter will be rather 
complicated due to the new vertices introduced by the gradient flow~\cite{Monahan:2017talk}. With larger statistics and the momentum-smearing technique, which allows high momentum with small statistical errors~\cite{Bali:2016lva}, the uncertainty of lattice simulations will soon be dominated by the renormalization, which, therefore, need to be properly addressed.

The renormalization of the quasi-PDF has been closely studied from the perturbative point of view~\cite{Xiong:2013bka,Ma:2014jla,Ji:2015jwa,Ishikawa:2016znu,Chen:2016fxx,Constantinou:2017sej,Ji:2017oey,Ishikawa:2017faj}.
The bare quasi-PDF suffers from both logarithmic and linear UV divergences~\cite{Xiong:2013bka,Ma:2014jla}. The linear divergence originates from the self-energy of the spacelike Wilson line $W_z(z,0)$ and can be absorbed into an exponential factor $\exp(\delta m|z|)$ where $\delta m$ has a mass dimension~\cite{Dotsenko:1979wb,Craigie:1980qs,Dorn:1986dt}. This linear divergence is not affected when the Wilson line is inserted between two separated quark fields, so the exponential factor is capable of removing the same divergence in the quasi-PDF~\cite{Ishikawa:2016znu,Chen:2016fxx}. Since the remaining divergences are logarithmic and can be subtracted by a renormalization factor that only depends on the endpoints, the quasi-PDF {has been proven} to be multiplicatively renormalizable {for $\tilde{h}(z,P_z,\tilde{\mu})$ at any individual z}~{\cite{Ji:2017oey,Ishikawa:2017faj}}. 

This property makes it possible to carry out a nonperturbative renormalization of the quasi-PDF in the regularization-invariant momentum-subtraction scheme (RI/MOM)~\cite{Martinelli:1994ty} that has been widely used for quark operators on the lattice. In the RI/MOM scheme, the UV divergence in the quasi-PDF can be  removed to all orders in perturbation theory by the renormalization constant determined non-perturbatively, leaving the theoretical uncertainty to how precisely one can match the renormalized quasi-PDF onto the $\overline{\text{MS}}$-renormalized PDF. Since the RI/MOM scheme is regularization independent, the matching kernel can be calculated analytically in the continuum theory with dimensional regularization ($d=4-2\epsilon$), which is free of linear divergence. The one-loop result has already been obtained, and shows nicely convergent features for Eq.~\ref{eq:factorization}, compared to the matching in the transverse-momentum cutoff scheme~\cite{Stewart:2017tvs}.

 {In Ref. \cite{Constantinou:2017sej}, the additional mixing due to the chiral symmetry breaking on the lattice was first identified in the 1-loop perturbative calculation of $O_{\gamma_z}$, where $O_{\gamma_t}$ can be free of this problem. We will come back to this mixing effect in the following sections. Note that $O_{\gamma_t}$ is also a feasible choice for the quasi-PDF in LaMET as it belongs to the same universality class of $O_{\gamma_z}$~\cite{Hatta:2013gta}. In a recent new proposal~\cite{Radyushkin:2017cyf,Orginos:2017kos} $O_{\gamma_t}$ is used to define an Ioffe-time or pseudo distribution, which requires the same lattice setting and similar factorization formula to extract out the PDF~\cite{Ji:2017rah}.}

In this work, we present lattice results for the nonperturbatively renormalized quasi-PDF in RI/MOM scheme for the isovector unpolarized case~\footnote{While this paper was being finalized, another paper~\cite{Alexandrou:2017huk} on the nonperturbative renormalization of the quasi-PDF appeared, where the authors discuss a similar renormalization prescription.}, and match it to the $\overline{\text{MS}}$ PDF at one-loop order in perturbative QCD.
We demonstrate the procedure with the previously calculated lattice quasi-PDF~\cite{Lin:2014zya,Chen:2016utp} using
clover valence fermions on $N_f = 2 + 1 + 1$ (degenerate up/down, strange and charm) flavors of highly improved staggered quarks (HISQ)~\cite{Follana:2006rc} generated by MILC Collaboration~\cite{Bazavov:2012xda}
 with lattice spacing $a = 0.12$~fm, box size $L \approx 3$~fm and pion mass $m_\pi \approx 310$~MeV. 
The presentation of the paper is organized as follows:
In Sec.~\ref{sec:th}, we provide the theoretical setup of the RI/MOM renormalization and explain how to implement it on the lattice. In Sec.~\ref{sec:results}, we show the result of the renormalization factor, and use it to renormalize our previous quasi-PDF~\cite{Lin:2014zya,Chen:2016utp} obtained on the same lattice. We then match the renormalized quasi-PDF in the RI/MOM scheme to the PDF in $\overline{\text{MS}}$ scheme following the procedure elaborated in Ref.~\cite{Stewart:2017tvs}. In Sec.~\ref{sec:sum}, we summarize our results and discuss possible directions for further studies. 


\section{Renormalization of Wilson-Link Operators}
\label{sec:th}

For continuum QCD, the renormalization of nonlocal quark bilinear
operators has been discussed since the 1980s
~\cite{Dotsenko:1979wb, Arefeva:1980zd, Craigie:1980qs, Dorn:1986dt},
and the multiplicative renormalizability of the operator has been proven~\cite{Ji:2017oey,Ishikawa:2017faj}. 
Recent studies based on one- and two-loop perturbative analysis~\cite{Ji:2015jwa,Ishikawa:2016znu,Chen:2016fxx} also indicate that this property might be valid to all orders.
Under this assumption, operator mixing does not appear for nonsinglet
operators in renormalization. 
Here, we address the situation in the lattice case, when certain symmetries are broken.

\subsection{Operator Mixing}

On the Euclidean lattice, QCD is invariant under discrete symmetries, which include parity ${\cal P}$,
time reversal ${\cal T}$ and charge conjugation ${\cal C}$.
The parity and time-reversal operation are generalized into any direction
in the Euclidean space. Because there is no distinction between time
and spatial directions, we call the generalized parity and time-reversal 
operations ${\cal P}_{\mu}$ and ${\cal T}_{\mu}$, respectively.
We investigate the transformation properties of the nonlocal operator {$O_{\Gamma}(z)$}
and as some of the discrete transformation can flip the sign of $z$,
it is convenient to define the combinations {\cite{Ishikawa:2017}}
\begin{eqnarray}
  O_{\Gamma\pm}(z)=&&\frac{1}{2}\Big[\bar{\psi}(z)\Gamma W_z(z,0)\psi(0) \\
  && \quad \pm \bar{\psi}(0)\Gamma W_z(0,z)\psi(z)\Big]. \nonumber
\end{eqnarray}
The operator $O_{\Gamma\pm}(z)$
is Hermitian or anti-Hermitian, depending on $\Gamma$.
For $\Gamma=\gamma_z$, $O_{\gamma_z+(-)}(z)$
is anti-Hermitian (Hermitian).
The transformation properties of ${\cal C}$, ${\cal P}_{\mu}$ and ${\cal T}_{\mu}$
prohibit $O_{\gamma_z}(z)$ from mixing with other operators except for
$O_{\cal I}(z)$, where $\cal I$ is the identity matrix.
In the zero quark mass limit, we have chiral symmetry (a continuous symmetry), which eliminates the mixing
between $O_{\gamma_z}(z)$ and $O_{\cal I}(z)$.
Some lattice fermions, such as Wilson-type fermions, explicitly break chiral symmetry and
introduce a mixing between $O_{\gamma_z}(z)$ and $O_{\cal I}(z)$.
The situation for the other vector operators,
$\Gamma=\gamma_x, \gamma_y$ and $\gamma_t$, is different.
Discrete symmetries alone prohibit their mixing with other $\Gamma$'s
even if chiral symmetry is broken.

The same discussion can also be applied to pseudoscalar, axial vector,
and tensor operators {\cite{Ishikawa:2017}}. Our analysis is consistent with what was found in one-loop lattice perturbation theory~\cite{Constantinou:2017sej,Alexandrou:2017huk}.

Therefore, the renormalization of the nonlocal vector operators for lattice fermions without chiral symmetry can be schematically presented as
\begin{eqnarray}\label{ZZ}
\begin{pmatrix}
 O_{\gamma_z}(z) \\  O_{\cal I}(z)
\end{pmatrix}
&=&
\tilde{Z} \times 
\begin{pmatrix}
 O_{\gamma_z}^R(z) \\  O_{\cal I}^R(z)
\end{pmatrix}, \nn
\\
&=&
\begin{pmatrix}
Z_{11}(z) & Z_{12}(z)\\ Z_{21}(z) & Z_{22}(z)
\end{pmatrix}
\begin{pmatrix}
 O_{\gamma_z}^R(z) \\  O_{\cal I}^R(z)
\end{pmatrix},
\\
O_{\gamma_{i\not=z}}(z)
&=&
 Z_{V_i}^{-1}(z)  O_{\gamma_{i\not=z}}^R(z),
\end{eqnarray}
where all $Z$'s are complex functions.
For the diagonal elements,
$\re[Z_{11(22)}(z)]=\re[Z_{11(22)}(-z)]$ and
$\im[Z_{11(22)}(z)]=-\im[Z_{11(22)}(-z)]$.
For the off-diagonal ones,
$\re[Z_{12(21)}(z)]=-\re[Z_{12(21)}(-z)]$ and
$\rm[Z_{12(21)}(z)]=\im[Z_{12(21)}(-z)]$.

In the past, $\Gamma=\gamma_z$ has been chosen for the unpolarized quark distributions.
As we discussed above, the renormalization for this operator involves mixing with the scalar
operator, whose signal is generally worse in lattice simulations.
Alternatively, as pointed out in Ref.~\cite{Xiong:2013bka}, one can choose $\Gamma=\gamma_t$ instead of $\Gamma=\gamma_z$ to define the unpolarized quasi-PDF. This choice also approaches the normal PDF in the infinite-momentum limit and has the advantage of avoiding the mixing problem. 
However,
the matching kernel, which involves vectors in the $z$ and $t$ directions, becomes much more complicated in this case. Therefore, we leave it for future investigation, and concentrate in this work on $\Gamma=\gamma_z$ and {will estimate the mixing effect from scalar operator }nonperturbatively.

\subsection{Nonperturbative Renormalization of the $O_{\gamma_z}(z)$ Operator 
in the RI/MOM Scheme}

The renormalization matrix elements of Eq.~(\ref{ZZ}) will be computed on the lattice as the amputated Green's function of $O_{\Gamma}(z)$ in an off-shell quark state $|p\rangle$ under the Landau gauge condition,
\begin{eqnarray} \label{eq:wallsource}
	&&\Lambda(p,z,\Gamma)\nn\\
	&=&S(p)^{-1} \left\langle \sum_w {\gamma_5}S ^{\dagger}(p,w+zn){\gamma_5}\Gamma W_z(w+zn,w)S(p,w) \right\rangle\nn\\
	&&\cdot S(p)^{-1} \ ,
\end{eqnarray} 
where $n^\mu=(0,0,0,1)$ is the unit vector along the $z$ direction
and the summation is over all lattice sites $w$. The quark propagators are defined as
\begin{equation}
	S(p,x)=\sum_{y} e^{ipy}\langle \bar{\psi}(x) \psi(y)\rangle\ ,\ S(p)=\sum_{x} e^{-ipx}S(p,x)\,.
\end{equation}
and {two $\gamma_5$ are inserted to the both side of $S^{\dagger}(p,w+zn)$ in Eq.~(\ref{eq:wallsource}) to get the necessary propagator $\sum_{y} e^{-ipy}\langle \bar{\psi}(y) \psi(w+zn)\rangle$.} 

By imposing the RI/MOM renormalization condition,
\begin{align}\label{RIMOMrenormcond}
\frac{\textrm{Tr}[\slashed p\Lambda(p,z,\gamma_z)]^R}{\textrm{Tr}[\slashed p\Lambda(p,z,\gamma_z)_\text{tree}]}|_{p^2=\mu_R^2,\ p_z=P_z}& = 1, \nn\\
\frac{\textrm{Tr}[\Lambda(p,z,{\cal I})]^R}{\textrm{Tr}[\Lambda(p,z,{\cal I})_\text{tree}]}|_{p^2=\mu_R^2,\ p_z=P_z}&=1, \nn\\
\textrm{Tr}[[\slashed p\Lambda(p,z,{\cal I})]^R_{p^2=\mu_R^2,\ p_z=P_z} &= 0, \nn\\
\textrm{Tr}[\Lambda(p,z,\gamma_z)]^R_{p^2=\mu_R^2,\ p_z=P_z} &= 0,
\end{align}
where the superscript $R$ denotes a renormalized quantity, and $\mu_R$ is the renormalization scale. Note that the vertex functions are projected with $\tilde{\Gamma}=\slashed p/p_z$ to avoid the ambiguity arising from additional operator mixing in the off-shell matrix elements~\cite{Stewart:2017tvs}, and the prescription of equating the proton momentum  $P_z$ to the quark momentum  $p_z$ is used. The renormalization matrix $Z(z, p_z,a,\mu_R)$ with lattice spacing $a$ is inverse of $\tilde{Z}$ in Eq.~(\ref{ZZ},) which
can be extracted via
\begin{eqnarray} \label{projection}
&& Z(z,p_z,a,\mu_R) = \tilde{Z}^{-1}(z,p_z,a,\mu_R) \\
&&\back\back \tilde{Z}(z,p_z,a,\mu_R)\equiv\left(\begin{array}{cc}
Z_{11} & Z_{12}\\
Z_{21} & Z_{22}
\end{array}\right)(z,p_z,a,\mu_R)\nn\\
&=&\frac{1}{12 e^{-ip_zz}}\nn\\
&&\left(\begin{array}{cc}
\textrm{Tr}[\tilde{\Gamma}\Lambda(p,z,\gamma_z)] & \textrm{Tr}[ \tilde{\Gamma}\Lambda(p,z, {\cal I}) ] \\
\textrm{Tr}[ \Lambda(p,z,\gamma_z) ] & \textrm{Tr}[ \Lambda(p,z, {\cal I}) ]
\end{array}\right)_{p^2=\mu_R^2,\ p_z=P_z}.
\end{eqnarray} 
We drop the renormalization of the quark self energy, since it only contributes to the overall constant factor, which can eventually be determined by normalizing $\int q(x,\mu) dx$ to unity. {It is equivalent to normalize the vector charge to unity. Such a strategy have been used in many recent nucleon matrix elements calculations like Ref.~\cite{Bhattacharya:2016zcn,Green:2017keo,Berkowitz:2017gql}, and the origin of this idea can be traced to the original reference of the RI/MOM scheme, which used the discrete conserved vector current to define the renormalization constant of the quark self energy~\cite{Martinelli:1994ty}.}
 
Next, the renormalized proton matrix element of $O_{\gamma_z}^R(z)$ is computed by: 
\begin{eqnarray} 
\tilde{h}_R(z,P_z,\mu_R) &=& Z_{VV} \langle P| O_{\gamma_z}(z) | P \rangle + Z_{SV} \langle P| O_{\cal I}(z) | P \rangle \nn\\ \nn \\
Z_{VV} &=& \frac{1}{\Det(\tilde{Z})} Z_{22}(z,P_z,a,\mu_R)  \nn\\
Z_{SV} &=& -\frac{1}{\Det(\tilde{Z})} Z_{12}(z,P_z,a,\mu_R)
\end{eqnarray} 
where $\Det(\tilde{Z})$ is the determinant of the renormalization matrix $\tilde{Z}$. The $a$ dependence on
the right-hand side cancels up to discretization errors of order $O(aP_z, a\mu_R)$. 
The renormalized quasi-PDF $\tilde q_R(x,P_z,\mu_R)$ in the RI/MOM scheme can be obtained by a Fourier transform:
\begin{eqnarray} \label{Fourier}
\tilde q_R(x, P_z,\mu_R)=\int_{-\infty}^\infty \frac{dz}{2\pi}\ e^{ixP_zz} \tilde h_R(z,P_z,\mu_R).
\end{eqnarray} 

In the next section, we apply the RI/MOM scheme to the renormalization of quasi-PDF on the lattice, and eventually extract the $\overline{\text{MS}}$ PDF through a sequence of systematic corrections.

\section{Lattice Calculations}
\label{sec:results}

The results of our lattice calculations are presented in two parts: The first part is the nonperturbative renormalization constants in RI/MOM scheme, the second part is the result of the isovector unpolarized PDF. The bare quasi-PDF is renormalized using the renormalization in the first part, and then matched to the PDF using the one-loop matching formula after the power corrections in $P_z$ are applied. The final result is the isovector unpolarized PDF of the proton in the $\overline{\text{MS}}$ scheme.

\subsection{Renormalization Constants in the RI/MOM Scheme}
For the renormalization calculation, we used 33 configurations on the $L^3\times L_t=24^3\times 64$ lattice ~\cite{Bazavov:2012xda} used for the previous quasi-PDF calculation ~\cite{Lin:2014zya,Chen:2016utp}.
We connected the ends of the quasi-PDF operator to the sinks of the momentum-source quark propagators with $p=2\pi(3/L,2/L,3/L,8/L_t)$, which enables us to take the volume average of the operator position as in
Eq.~(\ref{eq:wallsource}), which improves the signal-to-noise ratio. 
This treatment allows us to access all the operators with different
Wilson-link lengths, although we must repeat the calculation for different
momenta. 

The $p_z=6 \pi/L$ computation used $p=2\pi(3/L,2/L,3/L,8/L_t)$ while the $p_z=4 \pi/L$ computation uses $p=2\pi(3/L,3/L,2/L,8/L_t)$. To {make the RI/MOM scheme $\mu_R^2=p^2$ to be the same}, we use the same $p$ for the latter case as the former one but change the operator to $\overline{\psi}(y)\gamma_y W_y(y,0)\psi(0)$. In both cases, $\mu_R^2=5.74\mbox{ GeV}^2$. {In the renormalization procedure, there is an arbitrariness to define the renormalization constant as long as it subtracts all the UV divergences. In our case, the renormalization constant should depend on how we choose $\mu_R$ and $p_z$, and $p_z=P_z$ is just one of the choices that we can make. This $P_z$ dependence shall be cancelled in the matching, as the RI/MOM scheme dependence must be cancelled in our matching up to $O(\alpha_s^2)$ and the the discretization effects.} {The discretization effect introduced by this choice can be eliminated eventually when the simulation with the same $p_z$ (in the physical unit) are repeated at smaller lattice spacings and the continuum extrapolation are applied on the renormalized results with different lattice spacings.}

\begin{figure}[htbp]
\includegraphics[width=.4\textwidth]{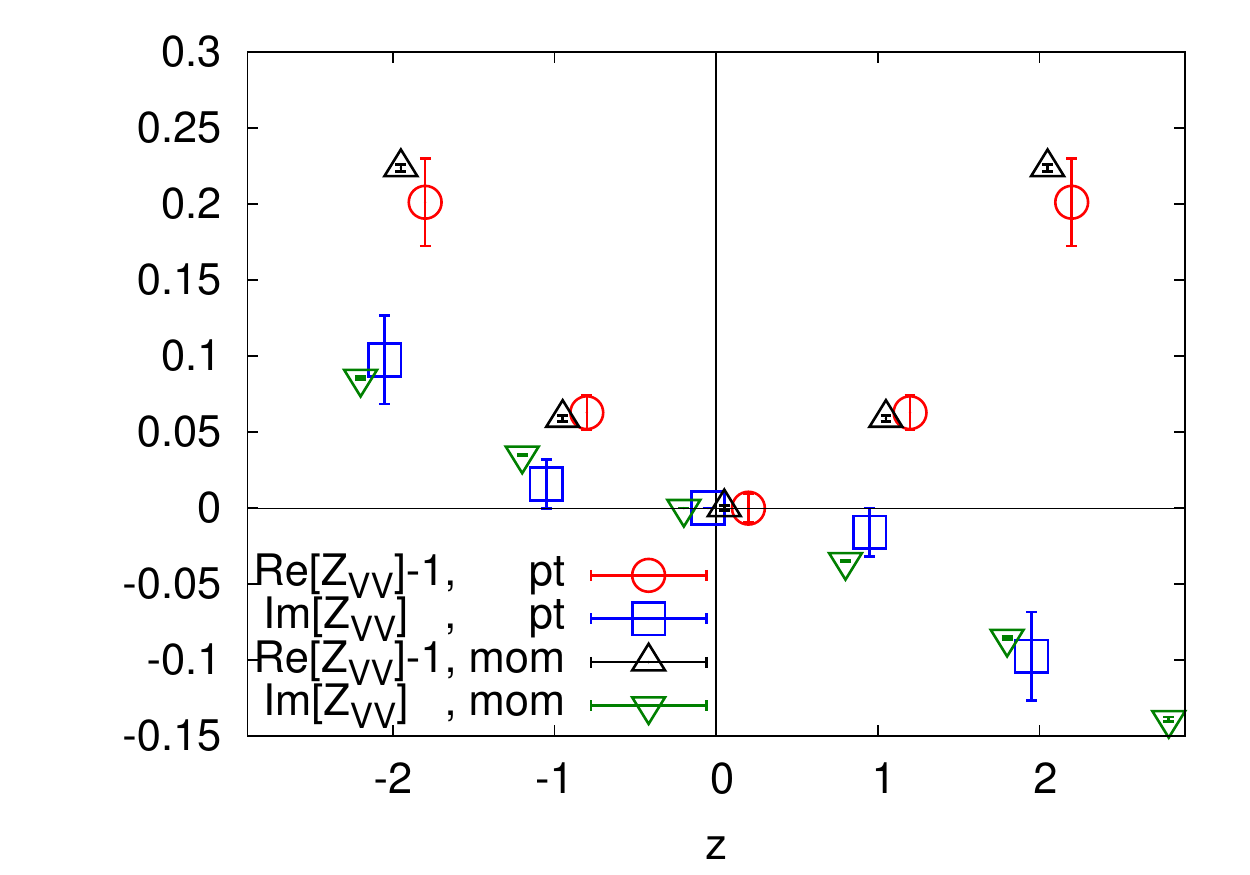}
\caption{Comparison between the renormalization constants obtained with the point source and the momentum source for $z\le 2$, taking the $p_z=6 \pi/L$ case as an example. The values are normalized by the central value of the renormalization constant at $z=0$ and the real parts are subtracted by unity for a better comparison. It is obvious that with the same configurations, the signal from the momentum source can be much better than that from the point source, while the values are consistent with each other.
} \label{fig:src_compare}
\end{figure}

A comparison of the signals between the point source and the momentum source for the $p_z=6 \pi/L$ case is given in Fig.~\ref{fig:src_compare}. It is obvious that with the same configurations, the signal with the momentum source can be much better than that with the point source, while the central values are consistent with each other.

\begin{figure}[tb]\label{hR}
\includegraphics[scale=0.7]{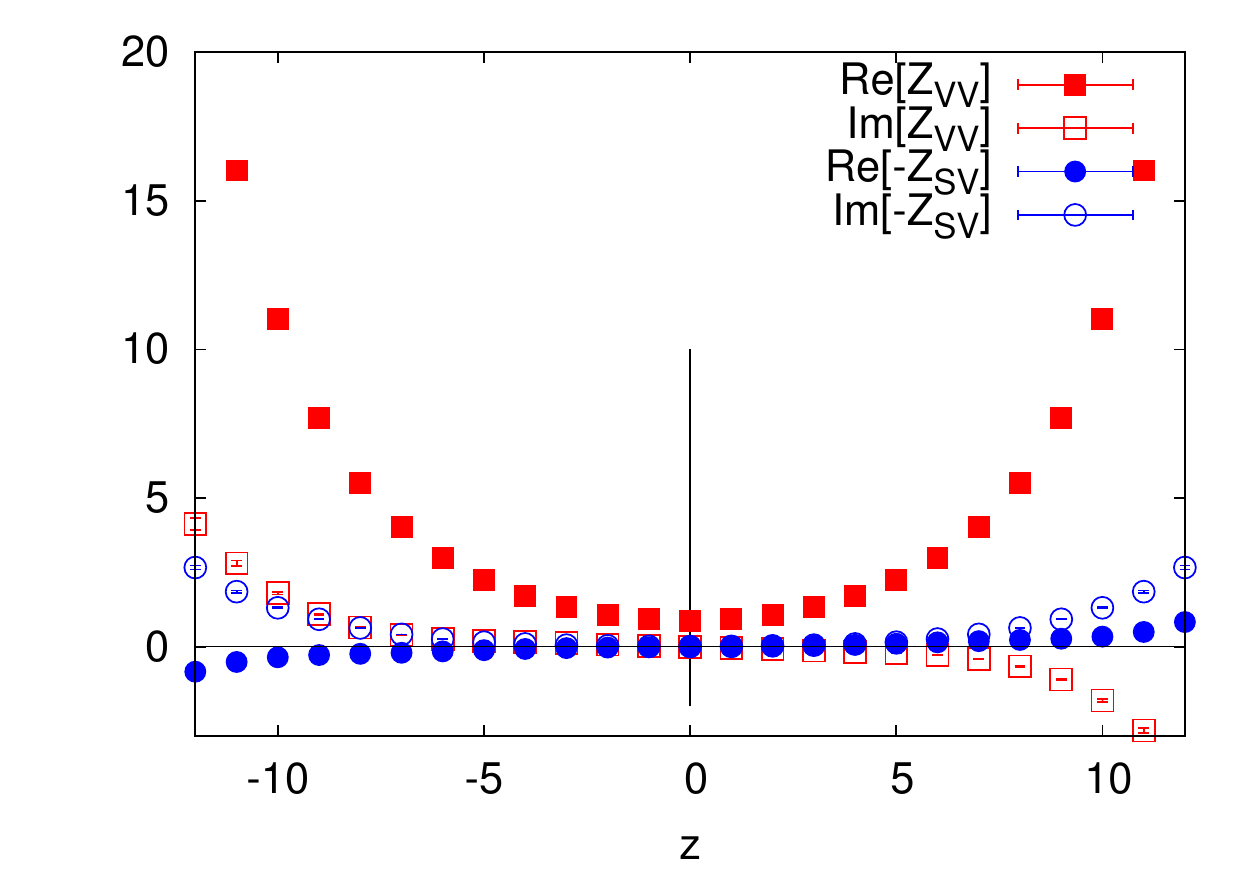}\\
\caption{The renormalization constant of the quasi-PDF operator $O_{\gamma_z}(z)$ (red boxes) and the mixing with the scalar quasi-PDF operator $O_{\cal I}(z)$ (blue dots) with the momentum along the Wilson link being $6\pi/L=1.29$~GeV and $\mu_R^2=p^2=5.74\mbox{ GeV}^2$. The size of the mixing coefficient is about an order of magnitude smaller than the renormalization factor in the large-$z$ region.
}\label{fig:Z:z}
\end{figure}

Fig.~\ref{fig:Z:z} shows both the renormalization factor and the mixing with the scalar quasi-PDF operator $O_{\cal I}(z)$ for the  $p_z=6 \pi/L$ case. {We would estimate the systematic uncertainty from ignoring the mixing by assuming $O_{\cal I}(z)\sim iO_{\gamma_z}(z)$ with the overall factor $i$ from the wick rotation, then $h_R(z)$ will not change at $z=0$, but it will change about $\sim8\%$ at $z=6$  and $\sim15\%$ at $z=12$ respectively, which is smaller than the present statistical uncertainties.}

\begin{figure*}[htbp]
\includegraphics[width=.4\textwidth]{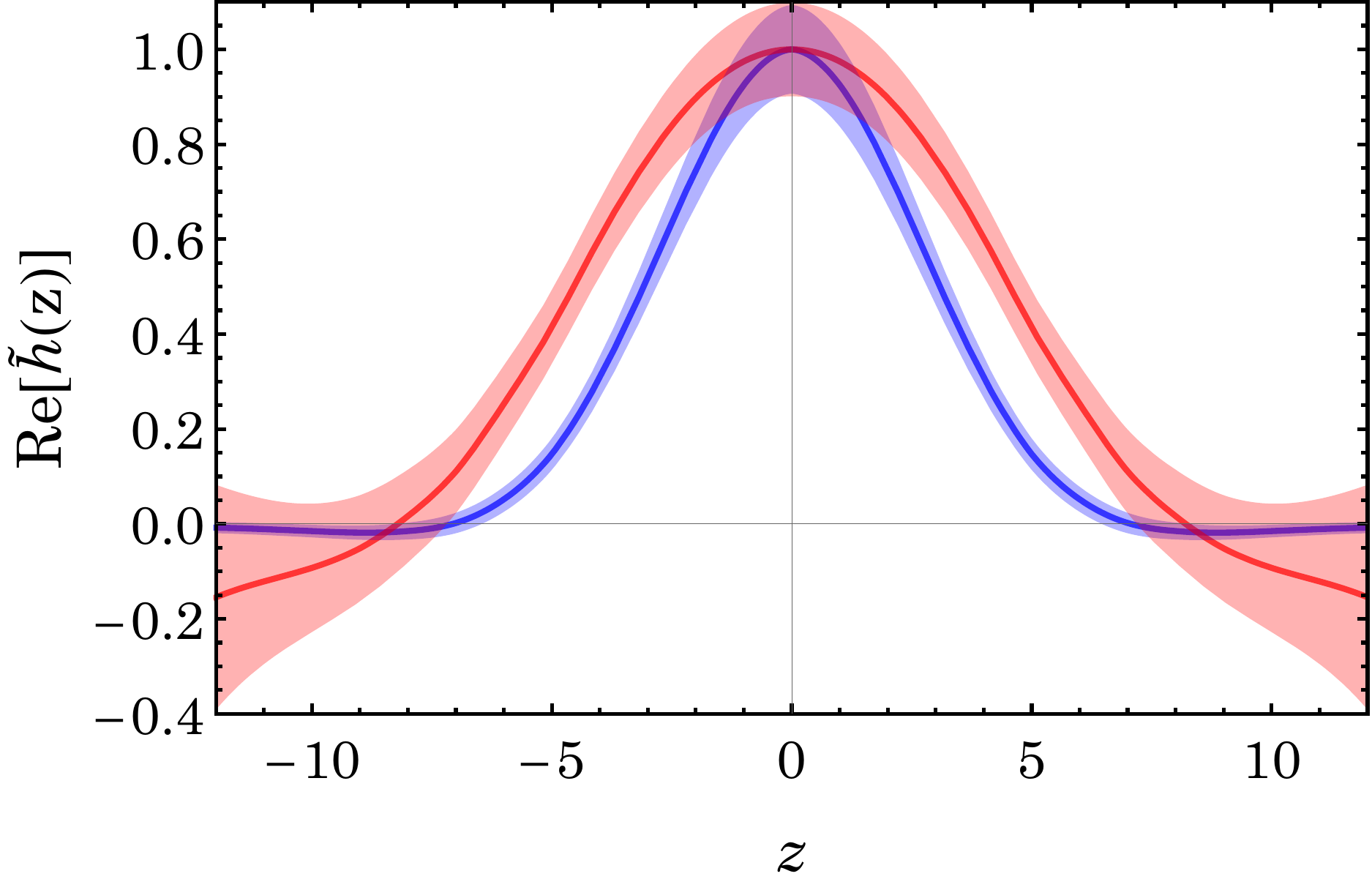}
\hspace{2em}
\includegraphics[width=.4\textwidth]{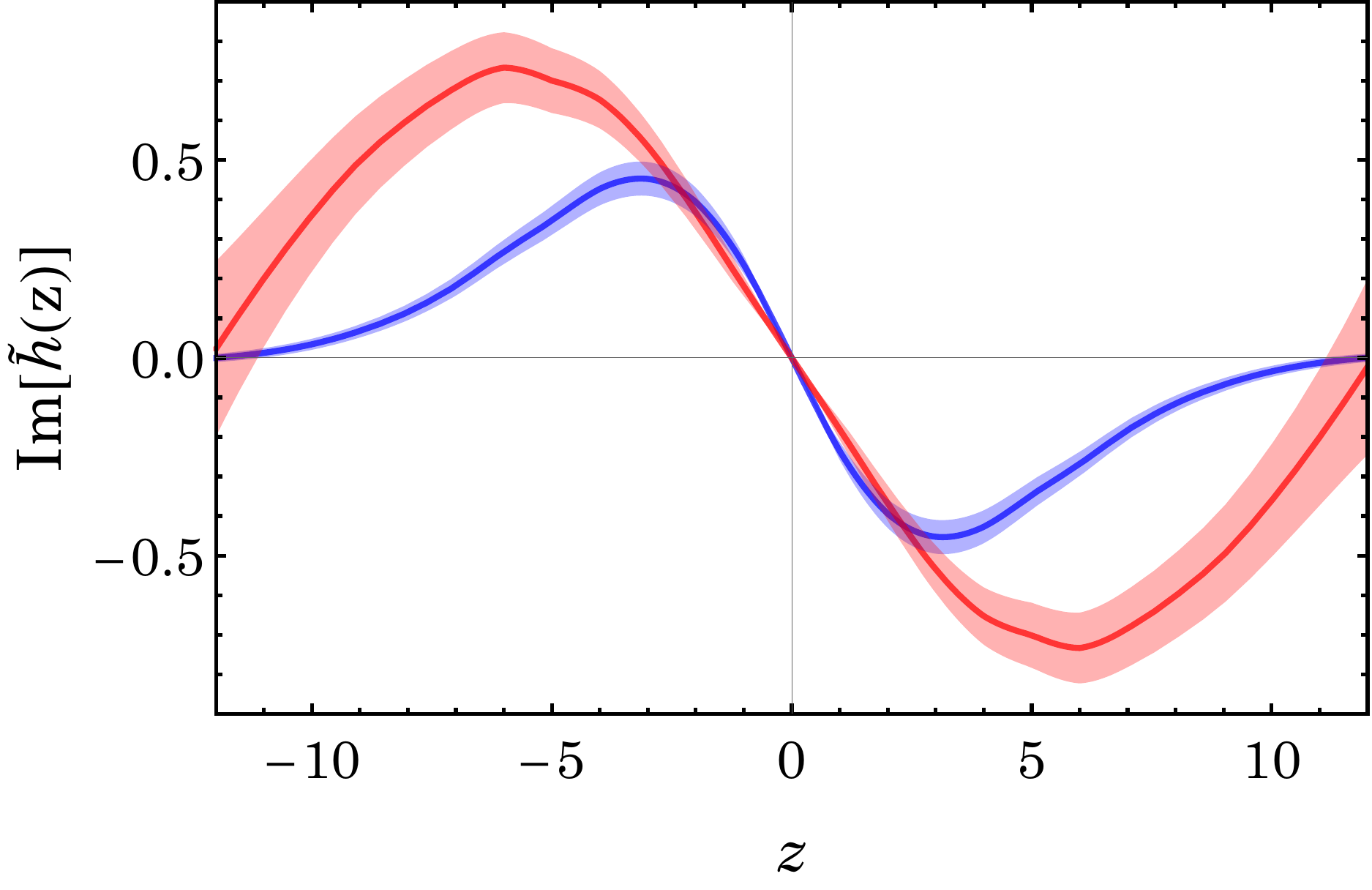}\\
\includegraphics[width=.4\textwidth]{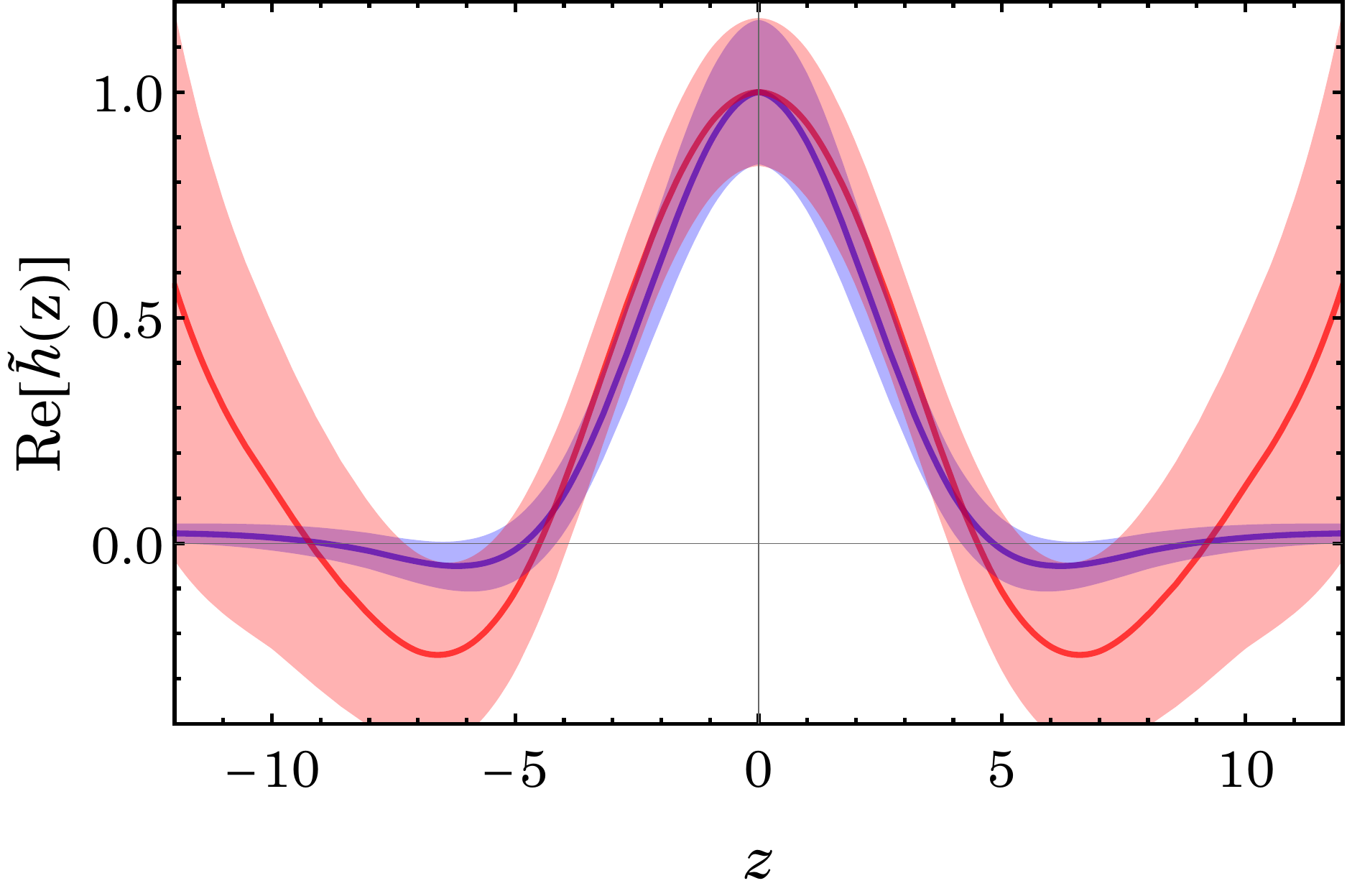}
\hspace{2em}
\includegraphics[width=.4\textwidth]{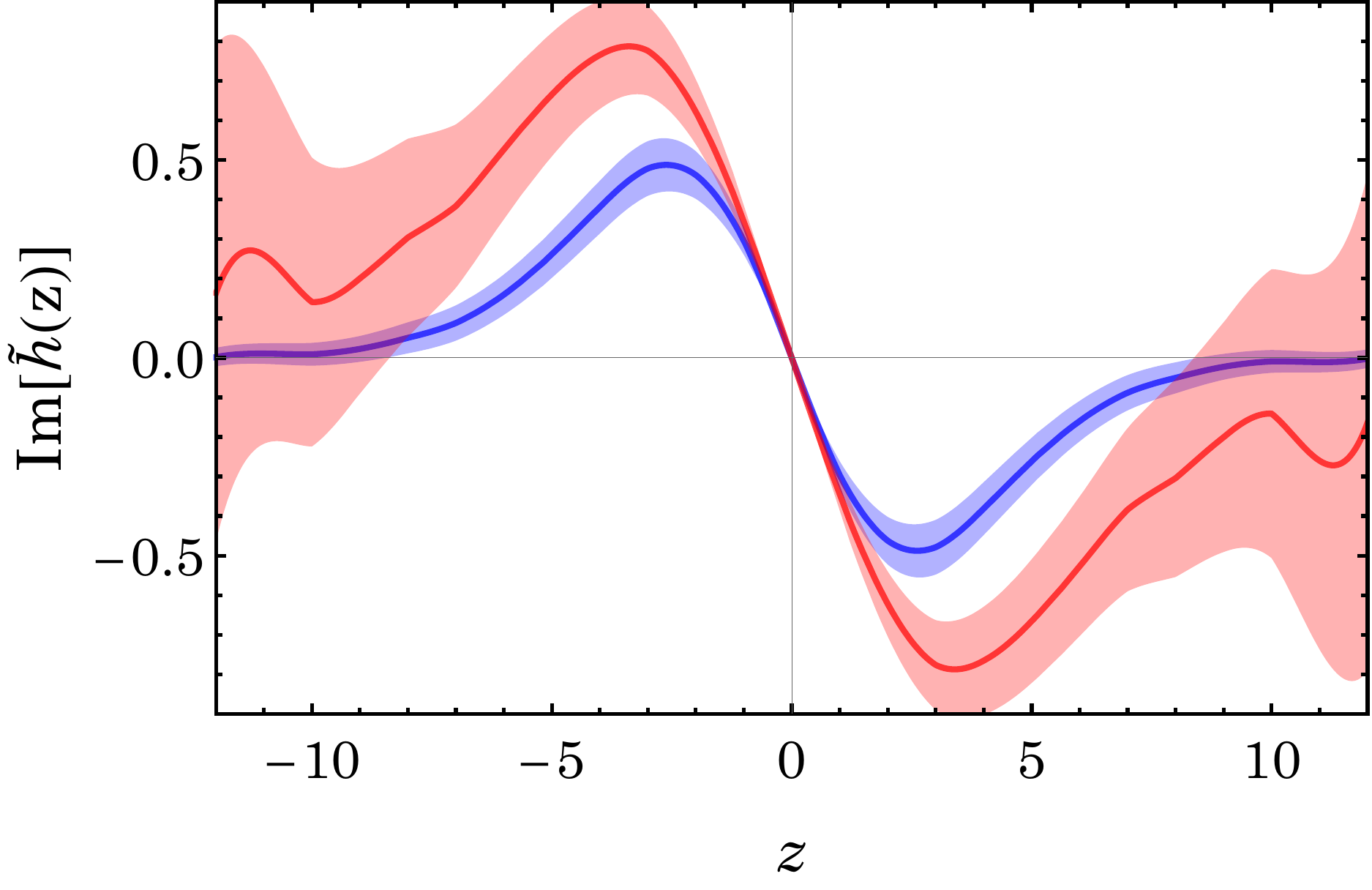}\\
\caption{The bare $ \tilde{h}_{\gamma_z}(z,P_z,\mu_R)$ (blue) and renormalized $ \tilde{h}_R(z,P_z,\mu_R)$ (red) for $P_z=4\pi/L$ (upper row) and $6\pi/L$ (lower row) with the renormalization scale $\mu_R=2.4$~GeV. The left and right panels show the real and imaginary parts, respectively.} \label{fig:ME}
\end{figure*}

In order to check the $\mu_R$ dependence, we repeated the renormalization calculations on two more momenta, $p=2\pi(5/L,2/L,3/L,10/L_t)$ and $p=2\pi(6/L,2/L,3/L,13/L_t)$. Since the matching from quasi-PDF to PDF will be processed in the space of the momentum fractions, we will check the $\mu_R$ dependence on the final distributions instead of that on the effective $\overline{\textrm{MS}}$ renormalization constants.

\subsection{From Quasi-PDF to PDF: Numerical Results and Discussion}

\begin{figure*}[tb]
\includegraphics[width=.6\textwidth]{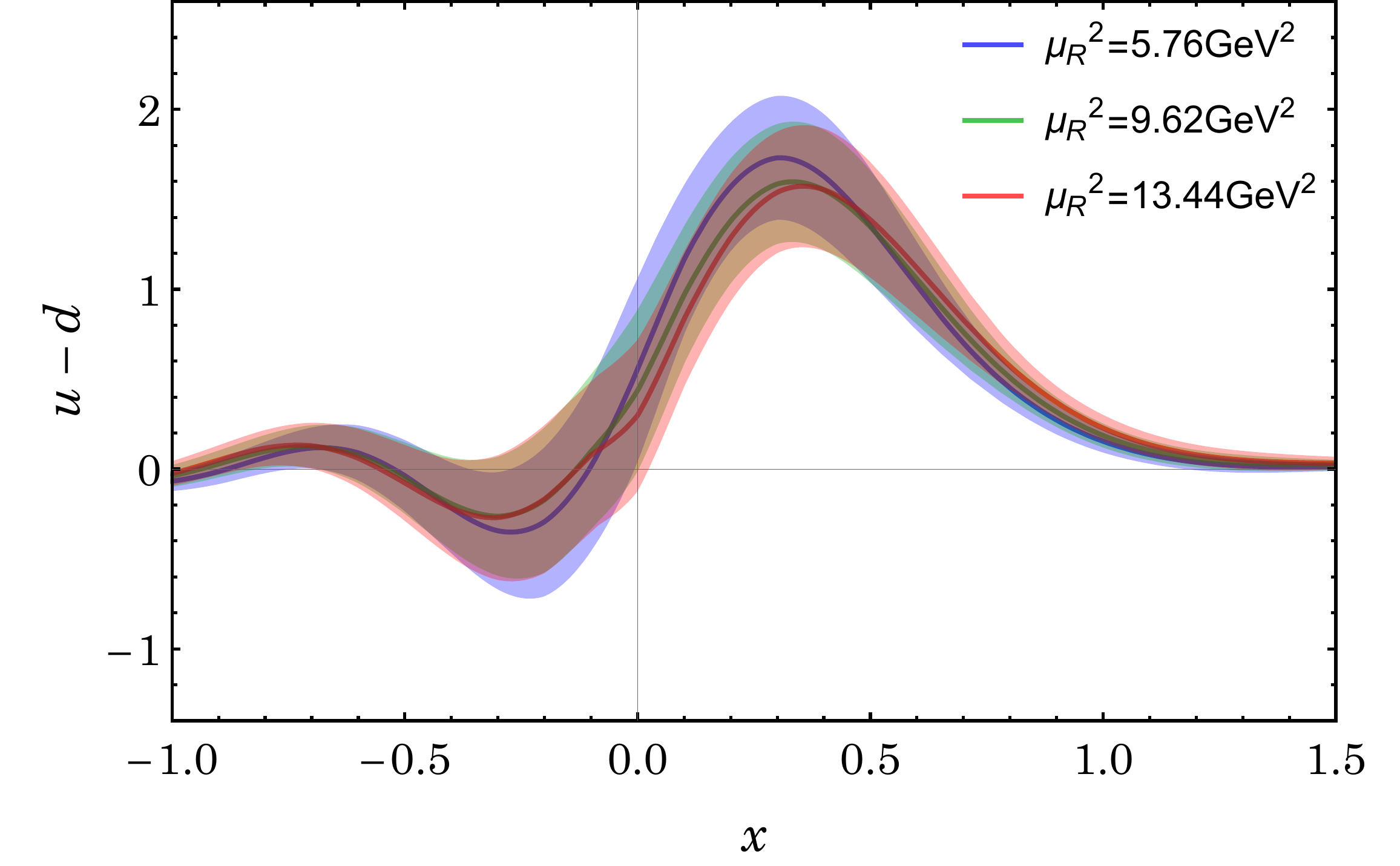}
\caption{{The renormalized unpolarized isovector quark distribution after one-loop matching and mass correction at the renormalization scale $\mu^2=5.76$~GeV$^2$ in the $\overline{\text{MS}}$ scheme, with $P_z=6\pi/L$ and three different RI/MOM renormalization scales $\mu_R$. Three distributions agree with each other within the statistical uncertainties, it shows the $\mu_R$ dependence is almost cancelled numerically. The negative-$x$ part is related to the antiquark distribution via $\bar{u}(x)-\bar{d}(x) = - u(-x)+ d(-x)$ for $x>0$.}
} \label{fig:mu_r_dep}
\end{figure*}

\begin{figure*}[tb]
\includegraphics[width=.6\textwidth]{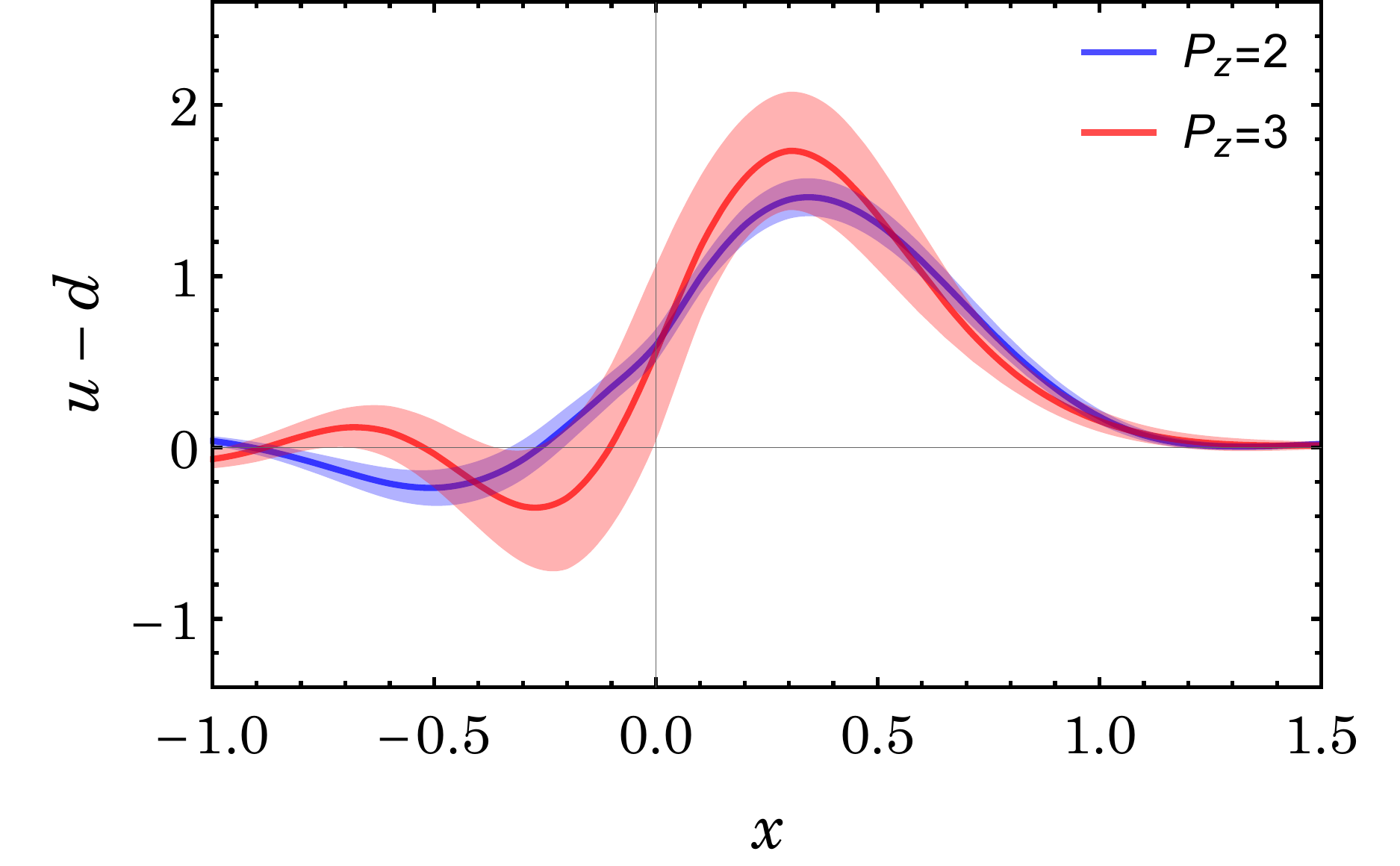}
\caption{The renormalized unpolarized isovector quark distribution after one-loop matching and mass correction at the renormalization scale $\mu^2=5.76$~GeV$^2$ {in the $\overline{\text{MS}}$ scheme}. The negative-$x$ part is related to the antiquark distribution via $\bar{u}(x)-\bar{d}(x) = - u(-x)+ d(-x)$ for $x>0$.
} \label{fig:final_pdf}
\end{figure*}

\begin{figure*}[htbp]
\includegraphics[width=.45\textwidth]{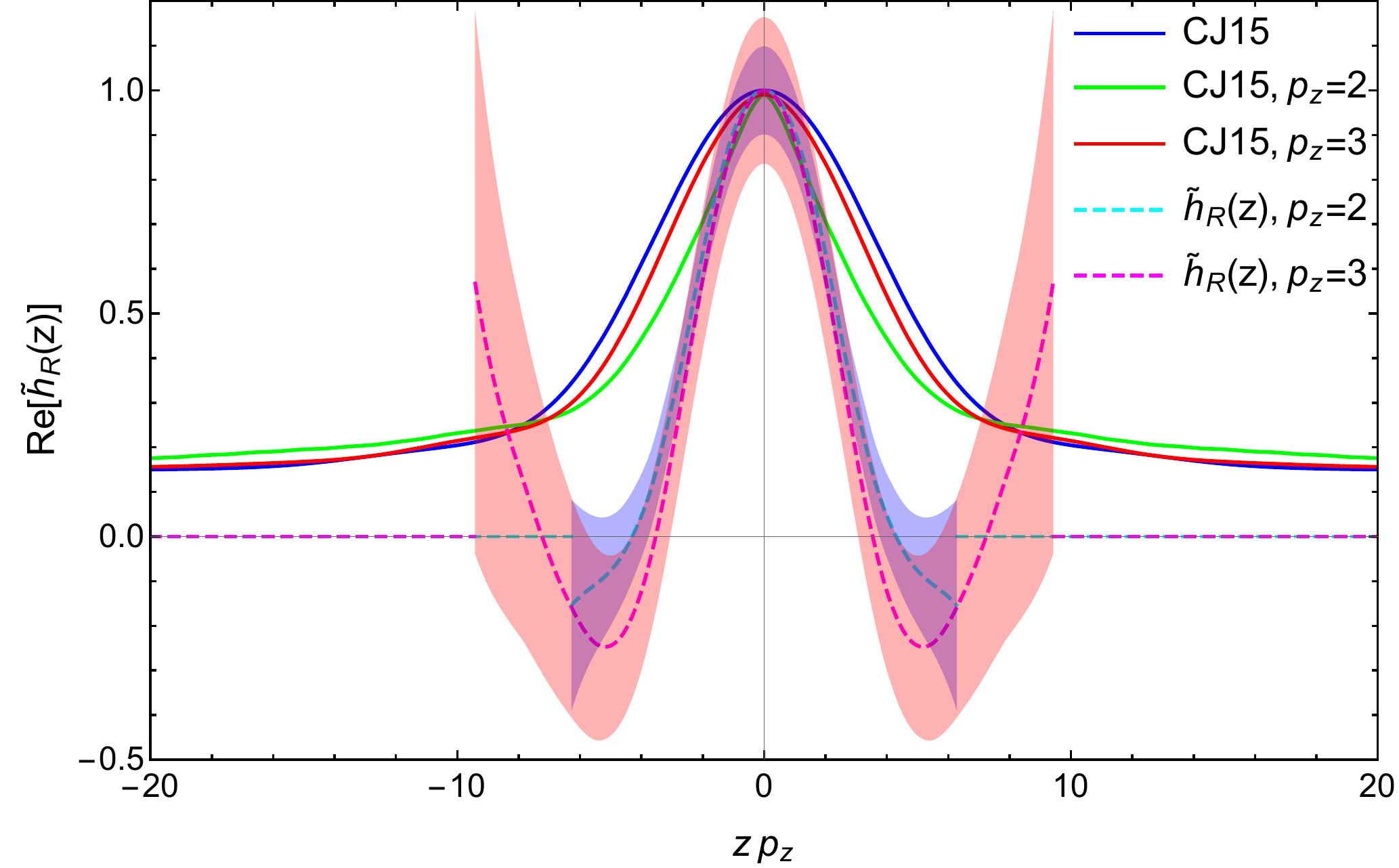}
\hspace{4em}
\includegraphics[width=.45\textwidth]{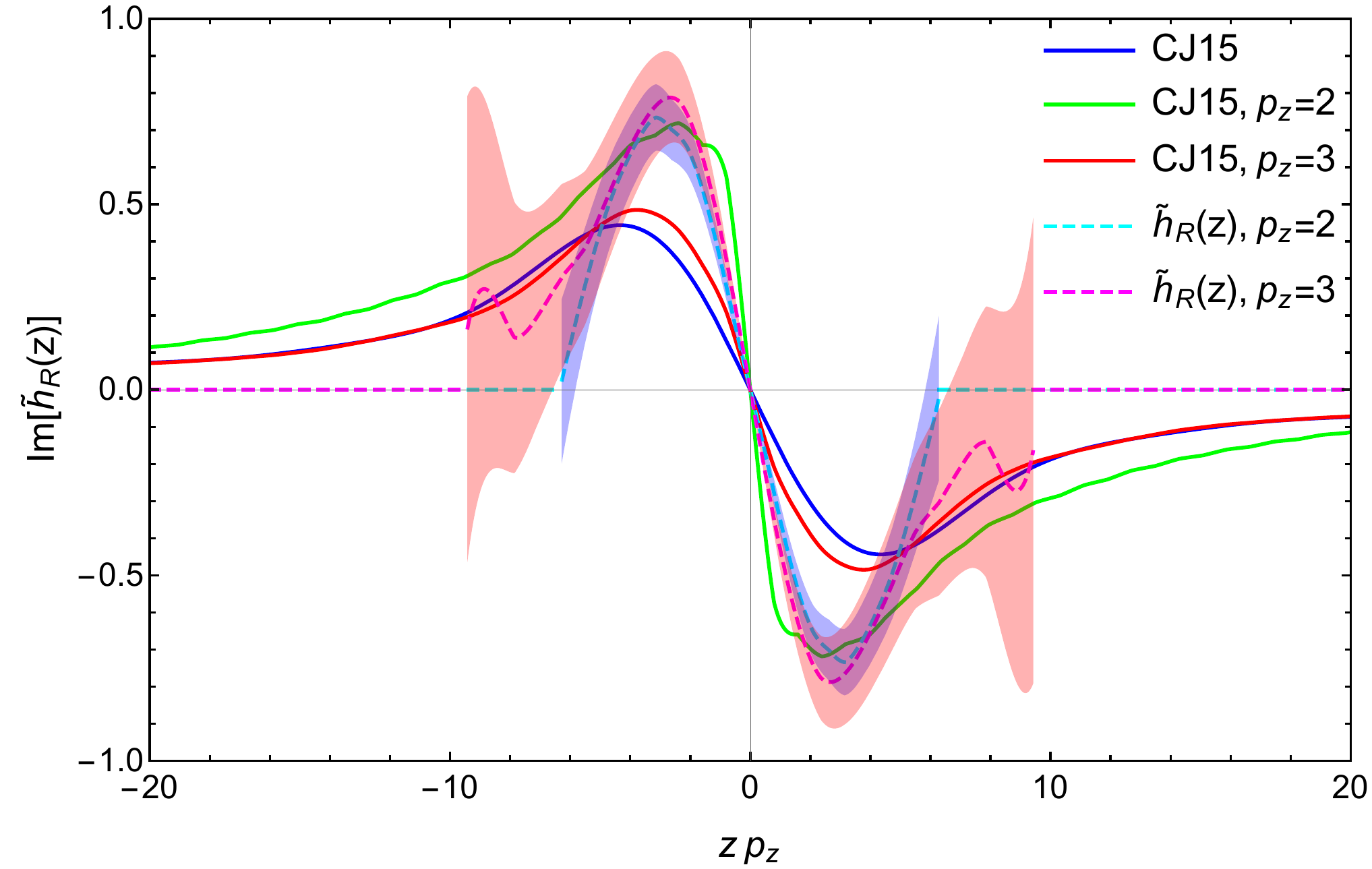}\\
\caption{Comparison of the {renormalized} function $ \tilde{h}_R(z)$ of this work (dashed lines) and from a Fourier transform of phenomenological PDFs {into the function of the Ioffe time $zP_z$}. {All the values are converted to $\mu^2=5.76$~GeV$^2$ in the $\overline{\text{MS}}$ scheme.} The solid lines are the Fourier transform of the corresponding CJ15 PDF (blue), after matching and mass corrections (green and red). 
The left and right panels show the real and imaginary parts, respectively.} \label{fig:hcomp}
\end{figure*}

In this subsection, we present our results for the unpolarized isovector quark distribution. We first calculate the time-independent, nonlocal correlator of a nucleon with finite $P_z$
\begin{align}
\label{eq:qlat}
h_{\Gamma}(z,\mu,P_z) =  
  \left\langle \vec{P} \right|
    \bar{\psi}(z) \Gamma \left( \prod_n U_z(n\hat{z})\right) \psi(0)
  \left| \vec{P} \right\rangle,
\end{align}
where $U_z$ is the gauge link pointing from $n\hat{z}$ to $(n+1)\hat{z}$, and $\vec{P}=(0,0,P_z)$ is the momentum of the nucleon. 
We calculate the bare lattice nucleon matrix elements $h_{\gamma_z}$ and $h_{\cal I}$
at $P_z=\{1,2,3\} 2\pi/L$, which are 0.43, 0.86 and 1.29~GeV, respectively. As observed in Refs.~\cite{Chen:2016utp,Zhang:2017bzy}, the correction terms for the smallest-momentum distribution is less well-behaved; thus, we drop it in the rest of this work. We then renormalize the bare matrix elements with the RI/MOM renormalization factors defined in the previous section:
\begin{align}
h_R\simeq Z_{VV} h_{\gamma_z},
\end{align}
where the mixing with $h_{{\cal I}}$ turns out to be numerically negligible because  $h_{\cal I}/h_{\gamma_z} \simeq M/P_z$ and $|Z_{SV}/Z_{VV}|  \ll 1$.
In Fig.~\ref{fig:ME}, we show the bare ($h_{\gamma_z}$) and renormalized ($h_R$) matrix elements for $P_z=\{2,3\} 2\pi/L$. In the renormalized matrix elements the mixing effect is temporarily ignored. We note that in both cases, the bare matrix elements vanish within error bands when the link length reaches 10{a}--12{a}. After renormalization, the error bands become much broader at large $z$ due to an exponential increase of the renormalization factor, and consistent with 0 within error bands.

Next, we  Fourier transform Eq.~(\ref{Fourier}) to convert 
the lattice matrix elements as functions of spatial link length $z$ into the quasi-PDF with $\mu_R$ the RI/MOM renormalization scale.
Then we take the one-loop RI/MOM-to-$\overline{\text{MS}}$ matching calculated in Ref.~\cite{Stewart:2017tvs} and mass corrections for the renormalized quasi-PDF. 
We invert Eq.~(\ref{eq:factorization}) to obtain the PDF in the $\overline{\text{MS}}$ scheme,
\begin{align} \label{eq:factorization3}
&q(x, \mu) =\tilde{q}_M(x, P_z,  \mu_R)\nn\\
&\quad-\frac{\alpha_s C_F}{2\pi}  \int_{-\infty}^{+\infty} \frac{dy}{|y|} C^{(1)}\left(\frac{x}{y}, \frac{\mu_R}{P_z},\frac{\mu}{|y|P_z}\right)\tilde{q}_{M}(y, P_z, \mu_R)\nn\\
&\quad+ {\cal O}\bigg(\frac{\Lambda_{\text{QCD}}^2}{P_z^2},\alpha_s^2 \bigg)\ ,
\end{align}
where $C^{(1)}$ has been computed in Ref.~\cite{Stewart:2017tvs},
	\begin{widetext}
	\begin{align} \label{eq:crimom}
	&C^{(1)}\left(\xi, {\mu_R\over p^z_R},{\mu\over p^z}\right)
	= C^{(1)}_{\tau=0}\left(\xi, {\mu_R\over p^z_R},{\mu\over p^z}\right)+(1-\tau) C_{\tau}^{(1)}\left(\xi, {\mu_R\over p^z_R},{\mu\over p^z}\right)\,,
	\end{align}
	where
	\begin{align} \label{eq:crimom1}
		&C^{(1)}_{\tau=0}\left(\xi, {\mu_R\over p^z_R},{\mu\over p^z}\right)\\
		&=\left\{\! \begin{array}{lc}
		\displaystyle \bigg[ {1+\xi^2 \over 1-\xi}\ln{\xi\over \xi-1} 
		-{2(1+\xi^2)-r_R\over (1-\xi)\sqrt{r_R-1}} \arctan {\sqrt{r_R-1}\over 2\xi-1}
		+{r_R\over 4\xi(\xi-1)+r_R} 
		\bigg]_\oplus  
		&  \xi>1\\[15pt]
		\bigg[ 
		\displaystyle{1+\xi^2\over 1-\xi} \ln{4 (p^z)^2 \over \mu^2 }+{1+\xi^2\over 1-\xi}\ln{\big[\xi(1-\xi)\big]}+(2-\xi) 
		-{2\arctan\sqrt{r_R-1}\over \sqrt{r_R-1}} \bigg\{ {1+\xi^2\over 1-\xi} -{r_R\over 2(1-\xi)} \bigg\}
		\bigg]_+    &0<\xi<1\\[15pt]
		\bigg[ \displaystyle{1+\xi^2\over 1-\xi}\ln{\xi-1\over \xi}+ {2\over \sqrt{r_R-1}}\left[{1+\xi^2\over 1-\xi} -{r_R\over 2(1-\xi)}\right]\arctan {\sqrt{r_R-1}\over 2\xi-1} - {r_R\over 4\xi(\xi-1)+r_R} \bigg]_\ominus  
		&\xi<0
		\end{array} \right. \,,
		\end{align}
		and
		\begin{align}
		C_{\tau}^{(1)}\Big(\xi, {\mu_R\over p^z_R},{\mu\over p^z}\Big)
		&= 
		\left\{ \begin{array}{lc} 
		\Bigg( \displaystyle {(1-2\xi)\over 2(1-\xi)} {r_R^2\over [r_R+4\xi(\xi-1)]^2}\Bigg)_\oplus  
		& \xi>1\\
		\displaystyle \biggl( -{(1-2\xi)\over 2(1-\xi)}\biggr)_+ 
		& 0<\xi<1\\
		\displaystyle \Bigg( -{(1-2\xi)\over 2(1-\xi)} {r_R^2\over [r_R+4\xi(\xi-1)]^2}\Bigg)_\ominus  
		& \xi<0 
		\end{array} \right.
		\,,
		\label{eq:landau}
		\end{align}
		with $r_R=\mu_R^2 /(p^z_R)^2$, and the Feynman and Landau gauges correspond to $\tau$=1 and 0 respectively. $p^z_R$ is the momentum used in the RI/MOM renormalization condition, whereas $p^z$ is the parton momentum for the matching. In our calculation, we choose $p^z_R=P^z$, and $p^z$ has to be $|y|P^z$ in the factorization theorem of Eq.~(\ref{eq:factorization3}). ``$\oplus$, $+$, and $\ominus$" denote plus functions whose defintions are given in~\cite{Stewart:2017tvs}.
	\end{widetext}

The $\tilde{q}_{M}(x, P_z, \mu_R)$ in the above equation is the quasi-PDF in the RI/MOM scheme with the nucleon mass correction removed~\cite{Lin:2014zya,Chen:2016utp},
\begin{align} \label{eq:mass_correction}
\tilde{q}_{M}(y)=&\sqrt{1+c}\sum_{n=0}^{\infty}\frac{\epsilon_c^n}{f_+}\big[(1+(-1)^n)\tilde{q}\big(\frac{f_+ x}{2\epsilon_c^n}\big)\nn\\
&+(1-(-1)^n)\tilde{q}\big(\frac{-f_+ x}{2\epsilon_c^n}\big)\big],
\end{align}
where $c=M_N^2/P_z^2$, $f_+=\sqrt{1+c}+1$ and $\epsilon_c\equiv c/f^2_+<1$ for any $P_z$. The remaining $\Lambda_{\text{QCD}}^2/P_z^2$ correction will be removed by a parametrization, as was done in Ref.~\cite{Chen:2016utp}. The $\mu_R$ dependence on the right-hand side should cancel modulo residual ${\cal O}(a^2\mu^2_R,\alpha_s^2)$ corrections. {The case with $P_z=6\pi/L$ and three difference $\mu_R$ is illustrated in Fig.~\ref{fig:mu_r_dep}. As in the figure, the residual ${\cal O}(a^2\mu^2_R,\alpha_s^2)$ dependence are smaller than the statistical uncertainties.}

The final results are shown in Fig.~\ref{fig:final_pdf}. 
 In contrast to the previous result in Ref.~\cite{Chen:2016utp}, the sea flavor asymmetry is hardly visible, mainly due to the rapid increase of the renormalization factor with distance, which amplifies the error. The peak in the positive-$x$ region is shifted slightly to the left. This is expected since the renormalization enhances the long-range correlation, and thereby enhancing the contribution in the small $x$ region when Fourier transformed to momentum space. After renormalization the unphysical dip near $x=0$ in the previous result also vanishes. This is because the linear divergence is removed and, therefore, the RI/MOM matching kernel has a smoother form than the matching used to relate bare PDFs. 

Another observation concerning our renormalized distribution is an oscillating behavior in negative-$x$ (antiquark) region, which is absent from the previous bare-PDF results. 
This is likely because the bare matrix element $\tilde{h}(z)$ decays very fast with the distance $z$, so the long-range correlation plays a less important role. However, the long-range correlation becomes more important in the renormalized distributions due to the exponential increase in the renormalization factor at large distance. {In such a case, the cut-off on $z$ will introduce large truncation errors by Fourier transforming the $\tilde{h}(z)$ into $\tilde{q}(x)$~\cite{Lin:2017ani}}.
To examine this hypothesis,  
{we apply the reverse matching and mass corrections procedure to the central values of one of the global fitted of PDF, ``CJ15'' (from the CTEQ-JLab collaboration~\cite{Accardi:2016qay}), to make a direct comparison with our renormalized function $h(z)$. Note that the PDF community fits $x q(x)$ and has larger uncertainty near $x=0$, so their $h(z)$ will also have large uncertainties at very large $z$.} The result is shown in Fig.~\ref{fig:hcomp}. 
{The renormalized $\tilde{h}(z)$ at 310-MeV pion mass and a relatively small lattice ($L$=2.9 fm) has a narrower peak around $zp_z$ = 0 and differs significantly from the Fourier transform of the CJ15 result at large values of $z p_z$. Further studies on removing the higher-twist contribution at large $z$ in the RI/MOM renormalization will be carried out in the future.}

Finally, we have several comments regarding our RI/MOM treatment. The first one is the possible gauge dependence induced by taking the external quarks off-shell in the nonperturbative renormalization. The gauge dependence should be canceled by the matching kernel, but the cancellation is not complete, since the kernel is only computed at one loop. It is encouraging that the one-loop matching effect is numerically small in Landau gauge that we employed. Whether the higher-loop contributions will remain small requires further study. 
The second one is treating $p_z$ of the off-shell quark the same as the proton $P_z$. Numerically, the renormalization factor is rather insensitive to $p_z$ and $\mu_R$, 
so we do not expect this treatment to cause a big error.  

\section{Summary}
\label{sec:sum}

We have carried out a nonperturbative renormalization of the quasi-PDF in the RI/MOM scheme in lattice QCD. Based on the renormalized quasi-PDF, we have updated the lattice result of the unpolarized isovector quark distribution from previous studies by some of the authors. 
The RI/MOM renormalization of the quasi-PDF is performed {for each individual $z$}, where it has been shown to be multiplicatively renormalizable. All the UV divergences, including the linear and logarithmic divergences, are subtracted nonperturbatively by the renormalization constant. Meanwhile, due to chiral symmetry breaking from {the lattice fermion action we used}, there is a mixing between the isovector quasi-PDF and a scalar operator. {The mixing is estimated to be a $\sim 10\%$ effect and is left for future investigation.}

{It is possible that an alternative process is needed in the large $z$ region. But the errors there are still too large to conclude. Nevertheless, there is an alternative renormalization method used in Ref.~\cite{Ji:2017oey,Green:2017xeu}  which is worth investigating. Also, two methods to reduce the weighting of the long range correlation are also discussed in Ref.~\cite{Lin:2017ani}. Clearly more efforts on this issue in the future are needed.}

Compared to the previous results on bare PDFs, our present result is free of the unphysical dip at $x=0$ due to the smooth matching kernel. However, we end up with a large uncertainty band that makes it difficult to evaluate whether an improvement has been achieved. The reason behind the large uncertainty band is that the RI/MOM renormalization constant which grows exponentially at large $z$ significantly amplifies the error in the nucleon matrix element of the quasi-PDF. Future work involving higher momentum, {larger volume, lighter pion mass,} and finer lattice spacing (such that the higher nucleon boosted momenta $P_z$ can be used without the additional $(P_z a)^n$ systematics) or other renormalization conditions may resolve some of the issues we see in this paper.

\vspace*{1cm}
\section*{Acknowledgments}
We thank the MILC Collaboration for sharing the lattices used to perform this study, the generation of those lattices used resources of Innovative and Novel Computational Impact on Theory and Experiment (INCITE) program by USQCD. The LQCD calculations were performed using the Chroma software suite~\cite{Edwards:2004sx}. 
Computations for this work were carried out in part on facilities of the
USQCD Collaboration, which are funded by the Office of Science of the
U.S. Department of Energy, on the National Energy Research Scientific Computing Center, and supported in part by Michigan State University through computational resources provided by the Institute for Cyber-Enabled Research. 
JHZ thanks Andreas Sch\"afer for helpful discussions.
This work was partially supported by 
the U.S. Department of Energy, 
Laboratory Directed Research and Development (LDRD) funding of BNL, under contract DE-SC0012704, a grant from National Science Foundation of China (No.~11405104), the SFB/TRR-55 grant "Hadron Physics from Lattice QCD", the MIT MISTI program, the Ministry of Science and Technology, Taiwan, under Grant Nos.~105-2112-M-002-017-MY3 and 105-2918-I-002 -003, the CASTS of NTU, and Kenda Foundation. 
The  work of JWC and YZ is supported in part by the U.S. Department of Energy, Office of Science, Office of Nuclear Physics, within the framework of the TMD Topical Collaboration. {TI is supported by Science and Technology Commission of Shanghai Municipality (Grants No. 16DZ2260200).} YZ is also been supported in part by the U.S. Department of Energy, Office of Science, Of- fice of Nuclear Physics, from DE-SC0011090.

\bibliography{latticepdf.bib}

\end{document}